\documentclass[a4paper]{jpconf}

\usepackage{graphicx}
\usepackage{amsmath}
\usepackage{hyperref}
\usepackage{bbm}
\usepackage{amssymb}
\usepackage[english]{babel}
\usepackage{lmodern}
\usepackage[T1]{fontenc}
\usepackage{wasysym}
\usepackage[normalem]{ulem}
\usepackage{color}
\usepackage[caption=false]{subfig}
\graphicspath{{Figs/}}

\usepackage{amssymb}

\begin{document}

\title{Studying and applying magnetic dressing with a Bell and Bloom magnetometer}

\author{Giuseppe Bevilacqua, Valerio Biancalana}
\address{Dept. of Information Engineering and Mathematics - DIISM, University of Siena -- Via Roma 56, 53100 Siena, Italy}

\author {Yordanka Dancheva}
\address{Dept. of Physical Sciences, Earth and Environment - DSFTA, University of Siena -- Via Roma 56, 53100 Siena, Italy; currently at Aerospazio Tecnologie Rapolano T. (SI) - Italy}

\author{Antonio Vigilante}
\address{Department of Physics and Astronomy, University College London, Gower Street, \\ London WC1E 6BT, United Kingdom}

\ead{valerio.biancalana@unisi.it}

\begin{abstract}
%\textcolor{red}{Atomic vapors are commonly investigated using a pump-probe scheme; in the suitable geometrical configurations and in the presence of a magnetic field atomic Larmor precession is detected by the probe, the observed precession frequency can be finely controlled using an appropriate alternating magnetic field, this phenomena is referred as atomic dressing.}
%\textcolor{blue}{Aspetto la rilettura di Dani e poi sistemo. Ok per una breve definizione di "dressing" qui all'inizio, ma la farei ancora più breve. Poi ci penso. Quanto a chiamare $x_0$ la posizione del sample, forse è meglio usare un altro simbolo ( $X_0, x_{sample} $?), perché, nelle eq 13-15, con $x_0$ ci si chiama la distanza dipolo-cella), in alternativa si può usare un altro simbolo ($d?, \Delta?$) nelle eq 13-15. Inoltre non so se nella caption va menzionata la telecamera, forse sì: aiuta a capire l'asse di destra. Ok per il resto delle revisioni}

The magnetic dressing phenomenon occurs when spins precessing in a  static field (holding field) are subject to an additional, strong, alternating field. It is usually studied when such extra field is homogeneous and oscillates in one direction.
We  study the  dynamics of  spins under  dressing condition in two unusual configurations. In the first instance, an inhomogeneous dressing field produces space dependent dressing phenomenon, which helps to operate the magnetometer in strongly inhomogeneous static field.
In the second instance, beside  the  usual configuration with static  and the  strong  orthogonal  oscillating magnetic   fields,   we  add  a secondary   oscillating   field, which is perpendicular  to  both.  The system shows novel and interesting features that are accurately explained and modelled theoretically. Possible applications of these novel features are briefly discussed.
\end{abstract}

\section{Introduction}
%\textcolor{blue}{
%Cose da fare con il dressing doppio che non vanno in questo articolo
%\begin{itemize}
%    \item SI può vedere alluminio sottile, e poi?
%    \item IDEA 2.0, lavorare con "gradienti" di dressing sulla J1
%    \item Misure inrinsecamente gradiometriche lungo x usando IDEA 2.0
%\end{itemize}
%}

The physics of a  precessing magnetization in a static   magnetic field is well-known since the studies performed by Joseph Larmor at the end of the 19th century. The phenomenology  of precessing  systems is enriched  when additional
time-dependent   fields are introduced.   In many applications
like  magnetic resonance  (MR) experiments  an  oscillating transverse
field (which is usually much weaker than the static one) is applied resonantly  to the Larmor one $\Omega_{\mathrm{L}}/2\pi$.  In  this  case the  resonant
condition helps in describing  the system dynamics in the 
rotating wave approximation: the  oscillating field is seen as a
superposition of two counter-rotating  components, one of which appears to
be  static in  the  reference  frame rotating  at  $\Omega_{\mathrm{L}}$ around  the
holding  field: it causes a slow  precession in  that rotating
frame.

The presence  of a transverse (off-resonant) alternating field  arouse interest also
in  the complementary  regime, when  its strength  exceeds  the static
one. This  condition is  technically unfeasible in  conventional   NMR at Tesla level, while  it is  more easily  accessible in
atomic physics experiments,  where the MR is commonly studied in much weaker fields,  e.g. at microtesla level.

The  seminal work  studying  precessing spins  subjected  to a  strong,
off-resonant, transverse field (\textit{dressing field}) dates back to
the late Sixties, when S.Haroche and co-workers
\cite{haroche_prl_70}    introduced   the   concept    of   magnetic
dressing.  That  contribution  started  a vivid  activity  around  the
subject \cite{Yabuzaki_jpsj_72, Kunitomo_pra_72,Ito_JPSJ_1994,Holthaus_pra_2001}
Recently, the magnetic  dressing was studied also with  cold atoms and
condensates  \cite{gerbier_pra_06,  Hofferberth_pra_07}. The  magnetic
dressing  of   atoms  offers  a  powerful  tool   in  quantum  control
experiments  \cite{gerbier_pra_06, previshko_prb_15} and high resolution
magnetometry  \cite{Swank_pra_18}.
%, and has been recently studied also in a regime  of  strong spin-exchange relaxation \cite{Hao_pra_19}.

Our  group  devoted research  activity  to  characterize the  magnetic
dressing  phenomenon occurring  when an atomic sample is subject to an  anharmonic dressing  field  \cite{bevilacqua_pra_12}.  More recently,  the
potential of applying a non-uniform  dressing field  to counteract the
effects  of static field  inhomogeneities was demonstrated \cite{biancalana_prappl_19}. This has important
  implications  in NMR  imaging  (MRI) in  the  ultra low  field  regime
\cite{biancalana_apl_19}. 

As a  general  feature,  the  mentioned works  studied  the  magnetic
dressing  in configurations  where the  precession around  the holding
field is modified by one time dependent field oriented along one
direction  in  the  plane   perpendicular  to  the  holding  one  (the
precession plane). 

In a more recent work \cite{bevilacqua_prl_20}, we studied theoretically and experimentally new features that emerge when the
design  of  the  dressing  field  uses  both  the  dimensions  of  the
precession  plane. We addressed the case when such time-dependent two-dimensional field is periodic in time. The periodicity is  obtained by superposing two perpendicular field components, which oscillate harmonically with  various amplitudes and with frequencies that are equal or in (small)  integer ratio $p$. 

The system is studied by means of a model that produces  analytical  results  thanks  to  a  perturbative
approach. This approach requires that one of the time-dependent components is much larger than both the static field and the other oscillating term. Despite its weakness, the latter (denominated \textit{tuning} field) plays an important role.

This paper is organized  as follows: in Sec.\ref{sec:model}, we describe
the model and derive  the expression for the effective precession frequency
in  terms  of the  dressing  parameters  (amplitudes, frequencies  and
relative phase of the dressing and tuning fields); in Sec.\ref{sec:idea}, we describe a practical application of magnetic dressing, which makes a Bell and Bloom atomic magnetometer suited to operate also in the presence of a strong field gradient.
In  Sec.\ref{sec:tuningdressing},  we  report experimental results obtained in the tuning-dressing configuration, and consider possible applications  where our findings may be of
interests.   An
appendix deepens some technical aspects of the  calculations on which
the model is based.

\section{Model \label{sec:model}}
The  evolution of  atomic magnetization  in an  external field  is commonly
described  using the  Bloch equations.  In this  work,  the considered
atomic  sample  is  a  buffered  Cs  cell at the core of  a  Bell  \&  Bloom
magnetometer  \cite{BellBloom1957}.  The   presence  of  a  \textit{weak}  pump
radiation synchronously  modulated at the  effective Larmor frequency
compensates the  \textit{weak}  relaxation phenomena  and enables the use  of a
model that  determines the steady-state solution on the basis  of the
simpler Larmor equation.

Consider the Larmor equation $\mathbf{\dot{M}} = \gamma \mathbf{B}
\times \mathbf{M}$ for the atomic magnetization
$\mathbf{M}$ in presence of a magnetic field of the form
\begin{equation}
  \label{eq:mf:form}
  \mathbf{B} = B_1 \cos (\Omega t) \mathbf{\hat{x}} +
               B_2 \cos ( p \Omega t + \phi) \mathbf{\hat{y}} +
               B_3 \mathbf{\hat{z}} 
\end{equation}
with $p$ an integer.  
We refer to $B_3$ as \textit{static} field, to $B_1$ as \textit{dressing} field and to $B_2$ as \textit{tuning} field, respectively.
Using the  adimensional time $ \tau = \Omega t $
one obtains explicitly ($\omega_i = \gamma B_i$)
\begin{equation}
  \label{eq:Larmor:adim}
  \frac{d\, \mathbf{M}}{d\, \tau} = 
  \left[ \frac{\omega_1}{\Omega} \cos (\tau)\, A_1 +
    \frac{\omega_2}{\Omega} \cos( p \tau + \phi) \, A_2 +
    \frac{\omega_3}{\Omega} A_3 
  \right] \mathbf{M},  
\end{equation}
where the three-dimensional matrices $A_i$ are antisymmetric with only
two  elements different  from zero  i.e. $(A_1)_{2,3}=  -(A_1)_{3,2} =
-1$.  $A_2$  and $A_3$  are obtained from  cyclic permutations  of the
indices: see eqs.\ref{eq:def:A_i} in the appendix. 

Assuming  that  $\xi=\omega_1/\Omega   \gg  \omega_2/\Omega  ,
\omega_3/\Omega$  the  perturbation theory let  factorize the
time  evolution operator  $U(\tau)$ ($\mathbf{M}(\tau)  \equiv U(\tau)
\mathbf{M}(0)$) in the interaction representation as
\begin{equation}
  \label{eq:U:repr:int}
  U(\tau) = \exp{\left[\xi\ \sin \tau \, A_1\right]}\, U_I(\tau) =
  \begin{pmatrix}
    1 &0 &0 \\
    0 &\cos \varphi & - \sin \varphi \\
    0 & \sin \varphi & \cos \varphi
  \end{pmatrix} U_I(\tau),
\end{equation}
where  $\varphi  =  \xi\ \sin  \tau  $,  obtaining  a
dynamical equation for $U_I(\tau)$: $d U_I /d \tau$ is given by
\begin{equation*}
  %\label{eq:UI}
   %\frac{d\, U_I}{d\, \tau}  = 
   \left[ 
    \left(      \frac{\omega_3}{\Omega}       \sin      \varphi      +
      \frac{\omega_2}{\Omega}  \cos  \varphi  \,  \cos(p\tau  +  \phi)
    \right) A_2  +
      \left( \frac{\omega_3}{\Omega} \cos \varphi -
        \frac{\omega_2}{\Omega}  \sin  \varphi  \,  \cos(p\tau  +  \phi)\right) A_3 
    \right] U_I 
    \equiv \epsilon A(\tau) U_I,
\end{equation*}
where  the parameter $\epsilon$  is  %a bookkeeping device  to  take  into  account  
used to identify 
the various  orders  of  perturbation
theory and to label the corresponding terms. 

The matrix $A(\tau)$ is periodic $A(\tau  + 2\pi) = A(\tau)$ so we can
use the Floquet theorem and write
\begin{equation}
  \label{eq:flo:magnus}
  U_I(\tau) = \e^{\Lambda(\tau)} \, \e^{\tau \,F}
\end{equation}
with $\Lambda(0)=0$  and $\Lambda(\tau+2\pi) =  \Lambda(\tau)$, and the
Floquet  matrix  $F$  is   time-independent.  Next,  we  expand  $U_I$
following the Floquet-Magnus \cite{magnus} expansion
\begin{equation}
  \label{eq:flo:magnus:expa}
  \Lambda  = \epsilon \Lambda_1 + \epsilon^2 \Lambda_2 + \cdots 
  \, , \, \, \, \, 
  F  = \epsilon F_1 + \epsilon^2 F_2 + \cdots 
\end{equation}
where the first terms are obtained from the relations
\begin{equation}
  \label{eq:F1:L1}
      F_1 = \frac{1}{2\pi} \int_0^{2\pi} A(\tau) \mathrm{d}\tau \, , \, \, \, \,
    \Lambda_1(\tau) =  \int_0^{\tau} A(\tau') \mathrm{d}\tau' -\tau
    F_1.
\end{equation}

Using the relation involving the Bessel functions $J_n$
\begin{equation}
  \label{eq:bessel}
  \e^{i  \,  z  \sin  \theta}  =  \sum_{n=-\infty}^{+\infty}  J_n(z)\,
  \e^{i\, n \, \theta}
\end{equation}
one finds
\begin{subequations}
  \label{eq:cos:sin:int}
  \begin{align}
    \int_0^{\tau} \cos( \varphi(\tau')) \mathrm{d} \tau' 
    & = J_0(\xi) \tau + f_1(\tau)  \\
    \int_0^{\tau} \sin( \varphi(\tau')) \mathrm{d} \tau' 
    & = f_2(\tau) \\
    \int_0^{\tau} \cos( \varphi(\tau')) \cos( p \tau' +\phi) \mathrm{d} \tau'
    & = \frac{1+(-1)^p}{2}\tau J_p(\xi) \cos\phi   + f_3(\tau) \\
    \int_0^{\tau} \sin( \varphi(\tau')) \cos( p \tau' +\phi) \mathrm{d} \tau'
    &= \frac{-1+(-1)^p}{2}\tau J_p(\xi) \sin\phi   + f_4(\tau).
  \end{align}
\end{subequations}
%where the  argument of the  Bessel functions is  $\xi$.
The functions $f_i(\tau)$ are reported in the appendix (see eqs.~\ref{eq:f_1:def}~-~\ref{eq:f_4:def}).

Using the eq.~\eqref{eq:cos:sin:int} one finds
\begin{equation}
  \label{eq:F1}
  F_1 =
  \begin{cases}
    \frac{\omega_3}{\Omega} J_0(\xi)  \, A_3 +  \frac{\omega_2}{\Omega} J_p(\xi)
    \cos \phi\, A_2 & p \, \mathrm{even} \\
    \left[ \frac{\omega_3}{\Omega} J_0(\xi) + \frac{\omega_2}{\Omega} J_p(\xi)
      \sin \phi \right] A_3 & p \,\mathrm{odd} 
  \end{cases}
\end{equation}
and 
\begin{equation}
  \label{eq:Lambda1}
    \Lambda_1 =  \left( \frac{\omega_3}{\Omega} f_2(\tau)  
      + \frac{\omega_2}{\Omega}f_3(\tau)\right) A_2 + 
           \left( \frac{\omega_2}{\Omega} f_1(\tau) - 
        \frac{\omega_2}{\Omega}f_4(\tau)\right) A_3 
\end{equation}
Using  the eq.~\eqref{eq:F1}  (see  appendix   for details)  we  can
calculate the dressed Larmor  frequency measured in the experiment for
the  component of the  magnetization parallel  to the dressing field:
\begin{equation}
  \label{eq:fin:W_L}
  \Omega_{\mathrm{L}} = 
  \begin{cases}
    \sqrt{\omega_3^2 J_0^2(\xi)  + \omega_2^2 J_p^2(\xi)\  \cos^2\phi} & p\;
    \mathrm{even}\\
    \left|\omega_3 J_0(\xi) + \omega_2 J_p(\xi)\ \sin \phi\right| & p\; \mathrm{odd}
  \end{cases}
\end{equation}
The eq.~\eqref{eq:fin:W_L} is at the focus of this paper. It includes also the well known and simpler case of a single dressing field: for zero values of the tuning field (i.e. $\omega_2=0$) it gives $\Omega_{\mathrm{L}}=|J_0(\xi) \omega_3|$. The latter is the instance discussed in Sec.\ref{sec:idea}, where only the dressing field is applied, with the peculiarity of a position-dependent  parameter: $\xi=\xi(x)$. The more general case, with the presence of both the dressing and the tuning fields, is discussed in Sec.\ref{sec:tuningdressing}.

\section{Atomic resonance enhancement with inhomogenous dressing}
\label{sec:idea}
An interesting magnetometric application of the dressing phenomenon is based on applying a position-dependent  dressing (and/or tuning) field to compensate atomic MR broadening caused by  inhomogeneities of the static field. 
Unless counteracted, the broadening induced by strongly inhomogeneous static field would severely deteriorate the performance of the magnetometer. 

Details of  the  magnetometric experimental apparatus  can be  found elsewhere \cite{biancalana_apb_16}. Beside the field control system described in ref. \cite{biancalana_apb_16}, here several additional coils enable the application of the time dependent (dressing and tuning) fields. In particular, the stronger dressing field is applied with a solenoidal coil surrounding the sensor and the weaker tuning field is applied by means of an Helmholtz pair.
Alternatively, when an inhomogenous dressing field is required, it is produced by a small source (a coil wound on a hollow-cylinder ferrite, with the laser beams passing across the hole) placed in the proximity of the atomic cell.

As discussed in Refs.\cite{biancalana_prappl_19, biancalana_apl_19}, the application of inhomogenous dressing field paves the way to detect NMR imaging signals \textit{in loco}, that is with an atomic sensor operating in the same place where the NMR sample is. In fact, the frequency encoding technique used in MRI requires the application of static gradients which would destroy the atomic resonance.

The basic principle of operation can be summarized as follows: let the NMR sample and the sensor atom be in a static field $(B_{3}(x))$  with an intensity dependent on the $x$ co-ordinate , thus $\omega_3=\omega_3(x)=\gamma (B_{3}(0)+Gx)$, where $G=\partial B_3/\partial x$ is the gradient applied to the frequency encoding purpose, which in our case is up to about 150 nT/cm. Over the centimetric size of the atomic sample, such a gradient would broaden the atomic MR  from its original few-Hz width up to kHz level, smashing it completely and preventing any magnetometric measurement.

\begin{figure}[t]
\centering
\includegraphics [angle=0, width=0.9 \columnwidth] {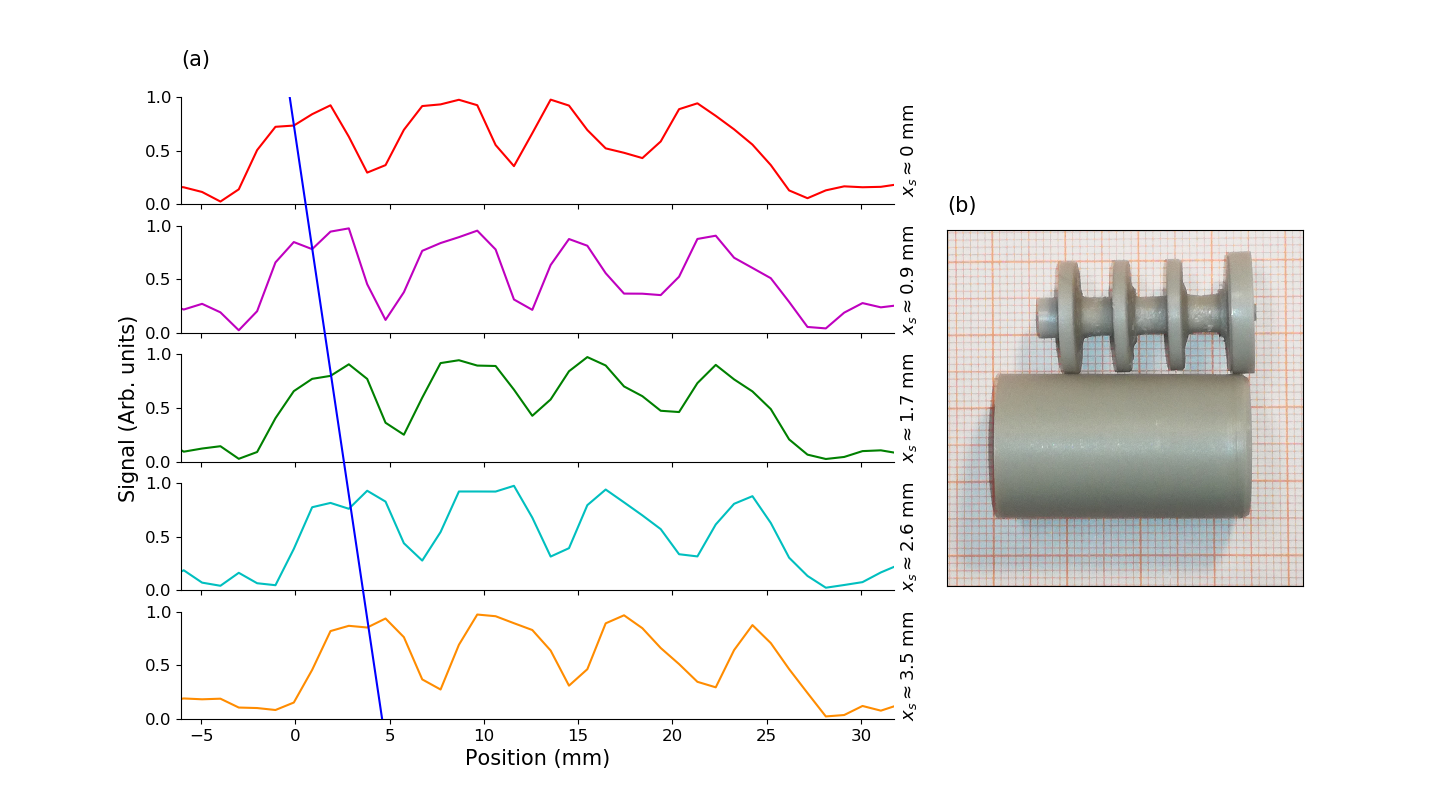}
\caption{1-D imaging of a structured (four disks) water sample contained in the plastic cartridge shown in the picture (b). The plots (a) represent MRI profiles, as obtained when the sample occupies four different positions $x_s$. Both the structure details and the positions (actual $x_s$ values are reported the right-side y-axes)  are determined with submillimetric precision.   }
\label{fig.submm}
\end{figure}

The narrow width of the atomic magnetic resonance can be restored by applying tailored tuning and/or dressing fields. We will consider here the case of single field dressing, where the gradient $G$  makes $\Omega_{\mathrm{L}}$ position-dependent according to

\begin{equation}
 \Omega_{\mathrm{L}}(x)=|J_0(\xi(x)) \omega_3(x)|,
\end{equation}
where we have considered a position dependent dressing field, making  the dressing parameter $\xi$ dependent on $x$, as well.

An appropriate inhomogeneity of $\xi$ makes possible to achieve the condition $\partial \Omega_L / \partial x =0$, under which the atomic resonance width is restored. It is worth noting that the nuclear precession is unaffected by the dressing field. The much smaller gyromagnetic factor, makes their dressing parameter vanishing and the effects of $B_1$ negligible: the MRI frequency encoding based on $G$ is preserved.

In this configuration, $B_1$ is inhomogeneous, and, in a dipole approximation, at a distance $x$ from the  cell center, the dressing field $B_x$ is
\begin{equation}
 B_x(x, t) =\frac{\mu_0}{2\pi}\frac{m(t)}{(x_0+x)^3} = B_1(x) \cos (\Omega t),
\end{equation}
where $\mu_0$  is the vacuum permittivity, $m(t)=m_0  \cos (\Omega t)$
is  the oscillating  dipole momentum, and $x_0$  is the  distance of  the dipole from the cell center.  
Taking now into account the dependence on $x$ of both the static and the dressing fields,  accordingly with the eq.\ref{eq:fin:W_L}, the dressed angular frequency results
\begin{equation}
\label{eq:omegaD}
\Omega_L(x)=\frac{\gamma_{\mathrm{Cs}}}{2 \pi}\left( B_0 + Gx\right) J_0 \left( \frac{\gamma_{\mathrm{Cs}} B_{1}(x)}{\Omega }\right),
\end{equation}
or, in a first-order Taylor approximation, 
\begin{equation*}
  \label{eq:eq:omega:svil}
  \begin{split}
  \Omega_{\mathrm{L}}(x)  = \Omega_{\mathrm{L}}(0) + \Omega_{\mathrm{L}}^{\prime}(0) \, x + O(x^2) %+ \frac{1}{2}  \nu_D^{\prime \prime}(0) x^2 + O(x^3) = \\
   \approx \frac{\gamma_{\mathrm{Cs}}}{2\pi} \left( B_0 J_0(\alpha) +  \left[ \frac{3 B_0 \alpha
  J_1(\alpha)}{x_0} + G  J_0(\alpha) \right] x \right ),\\
  \end{split}
\end{equation*}
where $\alpha = (\mu_0/2\pi) (\gamma_{\mathrm{Cs}} m_0)/(\Omega x_0^3) $. In conclusion, the condition for compensating the effect of the gradient $G$ is found  to be:
\begin{equation}
  \label{eq:def:grad}
   -3 \frac{B_0}{x_0} \frac{\alpha J_1(\alpha)}{J_0(\alpha)} = G. 
\end{equation}
This shows that with an appropriate choice of $\alpha$,  the width of the atomic magnetic resonance is substantially restored, as to recover the magnetometer performance to a level making it suited to detect weak MRI signals. 

The Figure \ref{fig.submm} shows MRI profiles obtained with this methodology. The frequency encoding is here obtained with a field gradient $G \approx 100$nT/cm. The presence of $G$ broadens the atomic MR from its  original 25 Hz up to about 800 Hz. Its width is then restored down to 35 Hz by inhomogeneous dressing.

\section{Tuning-dressing experiment}
\label{sec:tuningdressing}

%The dressed Larmor frequency  obtained in (\ref{eq:fin:W_L}) is tested using  an  optical  magnetometer  (OAM)  operating in  Bell  \&  Bloom configuration.  Details of  the  experimental apparatus  can be  found elsewhere \cite{biancalana_apb_16}.

The Eq.\ref{eq:fin:W_L} expresses the dependence of dressed angular frequency $\Omega_{\mathrm{L}}$ on the experimental parameters (strengths, relative phase and frequencies of the tuning and dressing fields). The predicted behaviour is verified using  the mentioned Bell and Bloom  magnetometer. 
%Three large size (180 cm), mutually orthogonal Helmholtz pairs are used to null the components  1 and 2 of the environmental  field and to reduce the component  3 at desired  level, and five  quadrupoles compensate the field gradients.
% the system
% contains a solenoidal coil wound around the atomic sensor to apply
% the component 1 of the dressing field and a small size (40 cm?)
% Helmholtz pair to apply the dressing field along the direction
% 2. The dressing coils are powered by two digital waveform generators
% in master-slave configuration. Having a shared clock, they produce
% signals that are independent but maintain a fixed and selectable
% relative phase.
%The dressing  fields are applied by means of two different coils. The stronger component $B_1$ is generated by  a solenoidal  coil wound around  the atomic  sensor, while the secondary dressing field $B_2$ is produced by a Helmholtz pair.  The coils are powered by two  waveform  generators sharing  one  internal  clock.  In  the experiment,  typical   values  of  the  three   field  components  are $B_1\approx 5-20 \mu$T  and $B_2, B_3\approx 1-4 \mu$T,  with the time dependent components $B_1$ and $B_2$ oscillating at $10-50$~kHz, with a precisely selectable relative phase.
Measurements  are  made  with 
$p=1,2,3$, and $\Omega_{\mathrm{L}}$ is recorded as
a function of $\omega_1$ and $\phi$.

%Figure \ref{fig.phases}(a)  shows the dependence  of $\Omega_{\mathrm{L}}$ on $\phi$. In the upper and bottom panels the variation follows the   characteristic  sine  profile,   the  different   amplitudes  are consistent   with   the   theoretical   values   of   $J_1(\xi)$   and $J_3(\xi)$, respectively. In the middle panel, in  accordance with the theoretical model, the variation follows the squared-cosine  profile and the amplitude is set by the value of $J_2(\xi)$.

\begin{figure}%[]

    \centering
    \subfloat[\label{fig:1a}]{
    \includegraphics [angle=0, width=.45 \columnwidth]{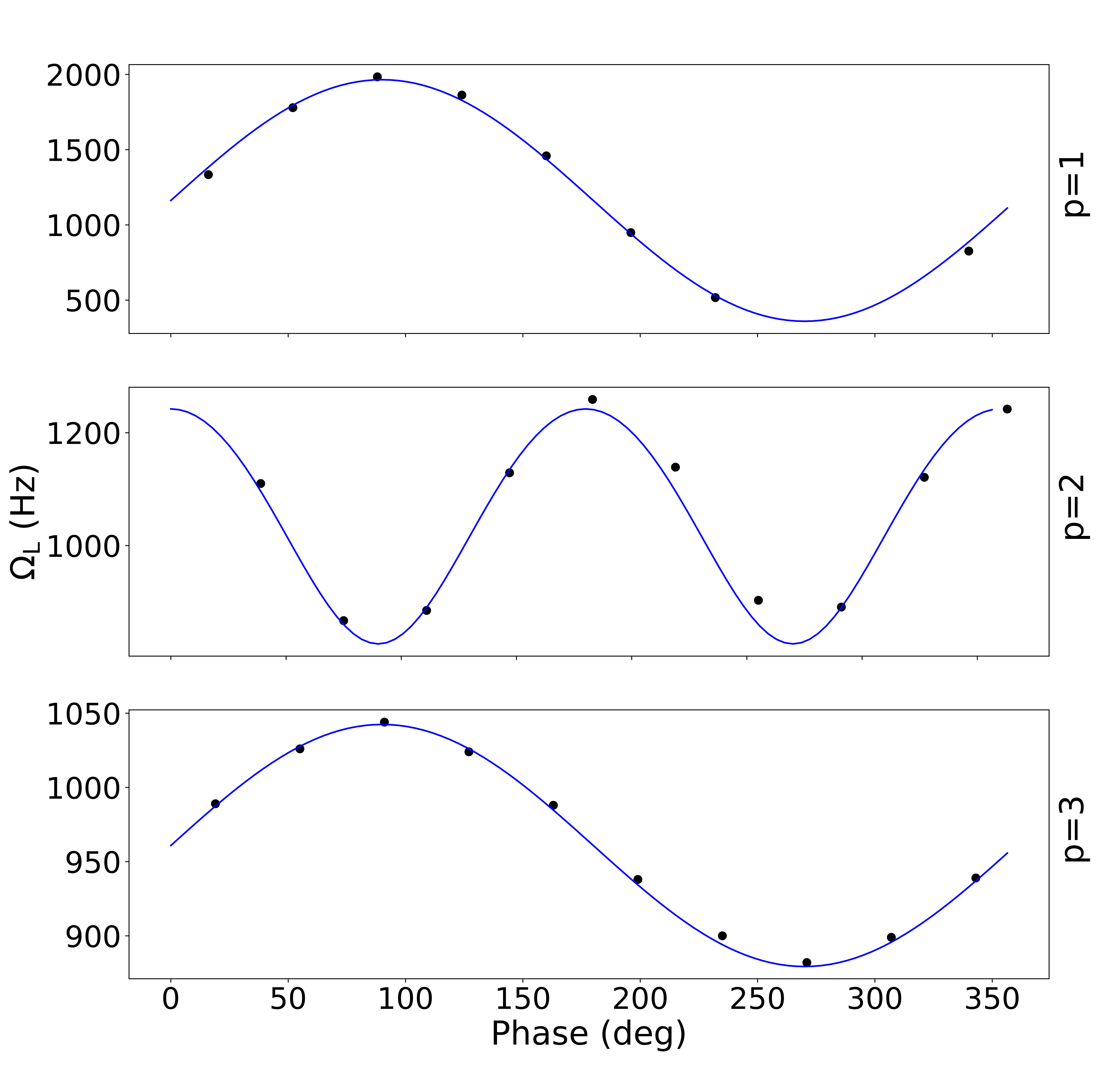}}
    \hfill
    \subfloat[\label{fig:1b}]{
    \includegraphics [angle=0, width=.45 \columnwidth]{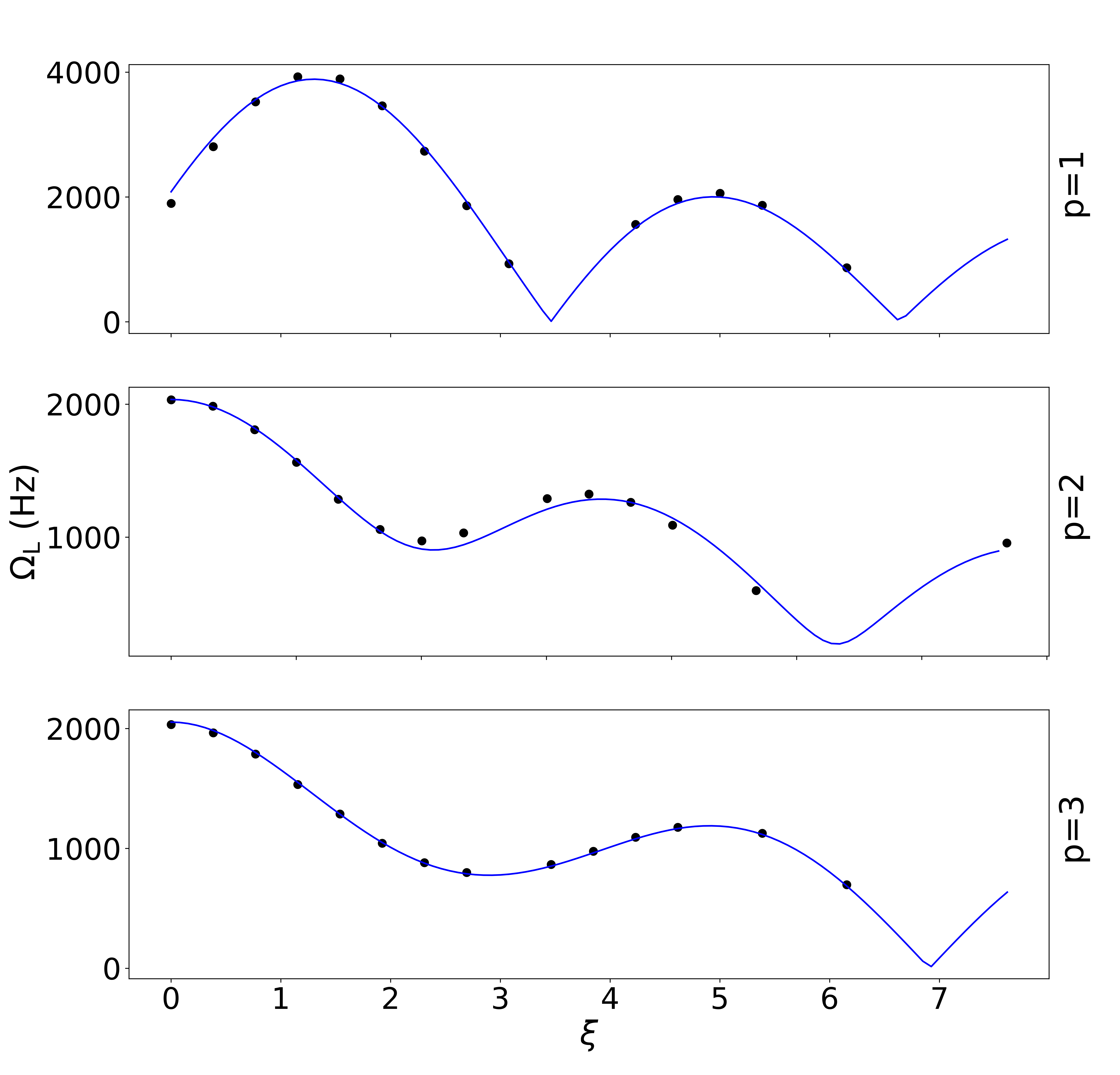}}
    \caption{(a) $\Omega_{\mathrm{L}}$ as a function of the relative  phase $\phi$ between  the fields  $B_1$ and  $B_2$. In black   dots  the  experimental  measurements,   in   blue  the theoretical predictions. Upper, middle and bottom panels refer respectively to the cases with $p=1, 2, 3$, in which $\xi$ has been fixed to $1.38,3.83 ,1.54$.
    (b)  $\Omega_{\mathrm{L}}$ as a function of the Bessel functions argument $\xi$.  In black dots the experimental   measurements, in  blue the theoretical  predictions. Upper, middle and bottom panels  refer respectively to the cases  with $p=1,2,3$. In the cases  with an odd  $p$ the phase  between $B_1$ and  $B_2$ is fixed to $\phi=\pi/2$, when $p$ is even the phase $\phi$ is set to zero.  In each  plot, $\xi$  changes varying  the value  of $B_1$, which ranges from zero to $\approx 15\ \mu$T. 
    }
  \label{fig.phases}
\end{figure}

% \begin{figure}[]
% \centering
% \includegraphics [angle=0, width= \columnwidth] {fig9.eps}
% \caption{(Color online) $\Omega_{\mathrm{L}}$ as a function of the Bessel functions argument $\xi$.  In black dots the experimental   measurements, in  blue the theoretical  predictions. Panels (a), (b) and (c)  refer respectively to the cases  with $p=1,2,3$. In the cases  with an odd  $p$ the phase  between $B_1$ and  $B_2$ is fixed to $\phi=\pi/2$, when $p$ is even the phase $\phi$ is set to      zero.  In each  plot, $\xi$  changes varying  the value  of $B_1$, which ranges from zero to $\approx 15\ \mu$T. }
% \label{fig.fields}
% \end{figure}

Figure \ref{fig.phases}  shows  comparisons of measured and calculated values  of $\Omega_{\mathrm{L}}$ as a  function of $\phi$ for given $\xi$ and  as a  function of $\xi$ for given $\phi$, respectively. The data shown in Figure \ref{fig.phases}(b) are measured with the $\phi$ values that maximize the tuning effect. i.e. is $\phi=\pi/2$ for the odd $p$ values and $\phi=0$ for the even one.  

There is  an excellent
accordance  between the experimental data  and the theoretical prediction. In particular, both the dependence of the relative phase $\phi$ and on the pertinent $J_p$ (with $p=1,2,3$) Bessel functions are perfectly verified by the experiment.  Minor deviations can be attributed to experimental imperfections, mainly to not-exact perpendicularity  of the applied fields $B_1$  and $B_2$.

The  developed  model  and   its  excellent  correspondence  with  the
experimental  results  demonstrates  the  possibility  to  enhance  the
control  of  spins evolution  by  means  of  the described  tuning-dressing field arrangement.

This  enhancement   opens  up  to   a  variety  of   new  experimental
configurations  in which  the new  set of  parameters ($B_2$,  $p$ and
$\phi$)   add   to   $B_1$    and   $\Omega$    to make new \textit{handles} available
to finely control the atomic magnetization dynamics.
Compared  to  the known  cases  of  harmonic \cite{haroche_prl_70}  or
anharmonic \cite{bevilacqua_pra_12} dressing field oscillating along one direction,  noticeably here the
dressed frequency $\Omega_\mathrm{L}$ (eq.\ref{eq:fin:W_L}) may exceed
$\omega_3$.

The tuning-dressing scheme makes  possible to chose  a parameter  set to
achieve an arbitrary $d \Omega_{\mathrm{L}}/d\xi$ in
conjunction   with   an   arbitrary   (to   some   extent)   value   of
$\Omega_{\mathrm{L}}$.  For instance, it is possible  to  produce a
condition of critical dressing (equalization of precession frequencies of different species) \cite{golub_pr_94,Swank_pra_18} with no
first-order  dependence  on $\xi$,  so  to  attenuate the  detrimental
effects caused by $B_1$ inhomogeneities, which 
constitute a severe limiting problem in high-resolution experiments \cite{Swank_pra_18}.

Similarly,  it is  possible to  fulfill the  condition of  a  large $d
\Omega_{\mathrm{L}}/d\xi$   avoiding  the   constraint  of   a  strong
$\Omega_{\mathrm{L}}$ reduction.  The latter may  help in applications
like that described in Sec.\ref{sec:idea},
where  $\Omega_{\mathrm{L}}$ is made  deliberately position-dependent
by  means of  a spatially inhomogeneous  $\xi$. In  addition, for  that  kind of
application,  it  is worth  noting  that  the  presented scheme  makes it
possible to  render $\Omega_{\mathrm{L}}$ space dependent  by means of
an inhomogeneity of the field $B_2$, which is of easier implementation
and control, being $B_2 \ll B_1$.

Other applications where the tuning-dressing scheme may offer
important  potentials is  suggested by  the dependence  on  $\phi$. As
recently  reported \cite{marmugi_apl_19},  an emerging  application of
highly sensitive magnetometers concerns  the detection of targets made
of weakly  conductive materials.  In that case,  the typical  setup is
based on a radio-frequency magnetometer, where the target modifies the
amplitude or the phase of a (resonant)  radio-frequency field driving
the magnetometer.  
Alternative setups  could be developed,
where the target modifies the field $B_2$, whose frequency is not required to match the atomic resonance. In this case, provided that
$J_p(\xi)$  is large (e.g. $\xi  \approx 1.84$  in the  case of  $p=1$), the
system  would have a  large response  to any  variation of  either the
amplitude (if  $\phi=\pm \pi/2$)  or the phase  (if $\phi=0,  \pi$) of
$B_2$ caused by eddy currents induced in the target.
%\textcolor{blue}{
%\begin{itemize}
%\item Nella  figura \ref{fig.04} la  differenza tra curva  arancione e
%  blu è  la calibrazione della bobina By.  A 27 kHz la  bobina ha una
%  caduta di tensione maggiore di circa il 5\%.
%\item Nella  figura \ref{fig.07} la  differenza tra curva  arancione e
%  verde è la calibrazione della bobina Bx, nel caso arancione vale la
%  calibrazione della tabella \ref{tab:bxcalibration01}, nel caso verde
%  vale la calibrazione della tabella \ref{tab:bxcalibration02}
%\end{itemize}
%}

%\input{conclusion.tex}

\appendix
\section{Appendix \label{sec:appendix}}

The explicit expressions for the matrices in the main text are
\begin{align}
  \label{eq:def:A_i}
  A_1 = 
  \begin{pmatrix}
    0 & 0 & 0 \\
    0 &0 & -1 \\
    0 & 1 &0
  \end{pmatrix} 
  \, , \,\, A_2 = 
  \begin{pmatrix}
    0 & 0 & 1 \\
    0 &0 & 0 \\
    -1 & 0 &0
  \end{pmatrix}
  \, , \,\, A_3 = 
  \begin{pmatrix}
    0 & 1 & 0 \\
    -1 &0 & 0 \\
    0 & 0 &0
  \end{pmatrix}
\end{align}

Defining $L_j  = i A_j$ and  taking the base  that diagonalizes $L_3$
one  obtains the  familiar  form for  the  angular momentum  operators
acting on the $| L=1, M_L \rangle $ states. This demonstrates that the
algebra generated  by the  $A_i$ matrices is  the same of  the quantum
angular  momentum operators.  Moreover, it  is straightforward to
demonstrate  that the  quantum mean  value $\langle  \gamma \mathbf{L}
\rangle$ satisfies the same classical Larmor equation.

The action of a general antisymmetric matrix  $W = a_1 A_1 + a_2 A_2 +
a_3 A_3$ on a given vector $\mathbf{v}$ is the cross-product
\begin{equation}
  \label{eq:cross}
  W \mathbf{v} = \mathbf{a} \times \mathbf{v} \qquad \mathbf{a} \equiv
  (a_1, a_2, a_3).
\end{equation}
Using the  Cayley-Hamilton theorem, the needed matrix    exponentials    can be analytically evaluated.   Let    write   $\mathbf{a}    =   \theta
\mathbf{\hat{a}}$ with  $\mathbf{\hat{a}} \cdot \mathbf{\hat{a}}  = 1$
then 
\begin{equation}
  \label{eq:expon}
  \e^{W} \mathbf{v} = \sum_{n=0}^{+\infty}\frac{W^n}{n!}\mathbf{v} = \mathbf{v} + 
  \sin \theta \, \mathbf{\hat{a}} \times \mathbf{v} + 
  (1 - \cos\theta) \, \mathbf{\hat{a}} \times ( \mathbf{\hat{a}} \times
  \mathbf{v} ). 
\end{equation}

The auxiliary $f_i$ functions introduced in the main text are defined as 
\begin{align}
  \label{eq:f_1:def}
  f_1(\tau)  &=  \sum_{n=1}^{\infty} \frac{J_{2  n}(\xi)}{n  }  \sin  ( 2  n
    \tau)\\
    f_2(\tau) &=  4 \sum_{n=0}^{\infty}  \frac{J_{2 n +  1}(\xi)}{ 2  n +1}
    \sin^2 ( (n+1/2)\tau)\\
    f_3(\tau) &= \Re ( g(\tau) )\\
    f_4(\tau) &= \Im ( g(\tau) )
    \label{eq:f_4:def}
\end{align}
where 
\begin{equation}
  \label{eq:g:def}
      g(\tau) =  \e^{i  \phi} \sum_{n \neq -p}\frac{J_n(\xi)}{i(n+p)} \left(
      \e^{i(n+p)\tau} -1\right) + \e^{- i \phi} \sum_{n \neq p}\frac{J_n(\xi)}{i(n - p)} \left(
        \e^{i(n -p)\tau} -1\right) 
  \end{equation}
%and all  the Bessel  functions have argument  $\omega_1/\Omega$.
These
functions have a  limited and oscillating behaviour and  are needed in
the evaluation  of $\e^{\Lambda_1(\tau)}$.  One can see  by inspection
that    $\e^{\Lambda_1}    \approx    \mathbbmss{1}$   is    a    good
approximation.  

The initial condition appropriate for the experiment is $\mathbf{M}(0)
\propto  (1,0,0)$ and  the quantity  monitored is  $M_x(\tau)$.  So applying
eq. \eqref{eq:expon}    to    evaluate    $\e^{\tau   F_1}$,    and    then eq.
\eqref{eq:U:repr:int}, one finds 
\begin{equation}
  \label{eq:Mx:fin}
  M_x(t) \propto \cos( \Omega_{\mathrm{L}} \; t).
\end{equation}

%%% Local Variables: 
%%% mode: latex
%%% TeX-master: "main"
%%% End: 

\section*{References}
% \begin{thebibliography}{9}
% \bibitem{iopartnum} IOP Publishing is to grateful Mark A Caprio, Center for Theoretical Physics, Yale University, for permission to include the {\tt iopart-num} \BibTeX package (version 2.0, December 21, 2006) with  this documentation. Updates and new releases of {\tt iopart-num} can be found on \verb"www.ctan.org" (CTAN). 
% \end{thebibliography}

\bibliographystyle{iopart-num}
\bibliography{bibliograph}
\end{document}